\documentstyle[12pt]{article}

\begin{document}
\thispagestyle{empty}
\begin{center}
\LARGE \tt \bf{ On 3D non-Riemannian geometry of twisted nematic liquid crystals}
\end{center}
\vspace{2.5cm}
\begin{center} {\large L.C. Garcia de Andrade\footnote{Departamento de
F\'{\i}sica Te\'{o}rica - Instituto de F\'{\i}sica - UERJ
Rua S\~{a}o Fco. Xavier 524, Rio de Janeiro, RJ
Maracan\~{a}, CEP:20550-003 , Brasil.
E-mail : GARCIA@SYMBCOMP.UERJ.BR}}
\end{center}
\vspace{2.0cm}
\begin{abstract}
A teleparallel geometrical description of the nematic phases of liquid crystals is proposed.In the case of the twisted geometry of nematics Cartan torsion is given by a spatial helical form which depends on the twist angle.
Geodesics of a test particle around the teleparallel liquid crystal are given.
\end{abstract}
\vspace{1.0cm}
\begin{center}
\large{PACS numbers : 0420,0450}
\end{center}
\newpage
\pagestyle{myheadings}
\markright{\underline{On non-Riemannian geometry of nematic liquid crystals}}
Several kinds of geometries have been used in the investigation of the elastic properties of 
solids.Among them Riemann-Cartan has been used by Katanaev and Volovich \cite{1} to show that 
it describes the defects (disclinations and dislocations) as solutions of 3D Einstein-Cartan 
equations.Earlier S.Amari \cite{2} has built a Finsler geometrical model of ferromagnetic 
substances.Riemannian geometry has also been used recently by myself \cite{3} to show that a 
Heisenberg ferromagnet yields a conical 2D solution of vacuum Einstein equations describing a 
monopole (topological defect)which could be match with Katanaev-Volovich solutions.This conical 
metric describes also cosmic strings in the Early Universe \cite{4}.Teleparallelism has also 
been applied with success \cite{5} to investigate a Bloch wall with spin density where the 
Cartan torsion \cite{6} is a distributional Dirac delta forming a sort of torsion wall.More 
recently Mullen  \cite{7} has also considered geometrical aspects of defects in semi-conductors 
in Josephson arrays.F.Moraes and C.Furtado \cite{8} have also recently used the Katanaev and 
Volovich ideas to investigate the graphite monolayer.In this letter we consider another 
application of teleparallelism \cite{9}, this time to nematic liquid crystals \cite{10},where the elastic properties of the nematic 
phases are investigated.Nematic liquid crystals can transmite torques and the application of 
torsion geometries will prove useful in the investigation of his geometrical and physical 
properties.Let us built the Riemann metric in analogy to Katanaev and Volovich as the 
deformation of the crystals $e_{ij}=n_{(j,i)}$ where this deformation represents the 
perturbation on the metric $g_{ij}$ given by     
\begin{equation}
g_{ij}={\delta}_{ij}+e_{ij}
\label{1}
\end{equation}
where ${\delta}_{ij}$ is the delta Kronecker symbol and the vector $n_{i}$ where ${(i=1,2,3)}$ is the director field of nematic liquid crystal.The teleparallel geometry was used by Einstein in 1928 \cite{10} to build the first teleparallel 4D geometrical unified theory of electromagnetism and gravitation.This theory has as one of his main carachteristics that the Riemann curvature tensor vanishes which implies that the anti-symmetric connection (torsion tensor) is given in terms of the metric by
\begin{equation}
{T}_{ijk}=g_{jk,i}-g_{ji,k}
\label{2}
\end{equation}
Here the comma denotes partial derivatives with respect to the lower index.Substitution of expression (\ref{1}) into (\ref{2}) yields the following torsion vector 
\begin{equation}
T_{k}=[{\nabla}^{2}n_{k}-{\delta}_{k}(div n)]
\label{3}
\end{equation}
therefore and the Weitzenb\"{o}ck condition for teleparallelism on the curvature Riemann tensor $R_{ijkl}({\Gamma})=0$ ,where ${\Gamma}$ is the Riemann-Cartan affine connection.Later on we shall make an application of this formula to a special case of nematic crystal.In the meantime let us compute the geodesics equations for the corresponding metric of the liquid crystal.The geodesics equations
\begin{equation}
\ddot{{x}^{i}}+{\Gamma}^{i}_{jk}{\dot{x}}^{j}{\dot{x}}^{k}=0
\label{4}
\end{equation}
and
\begin{equation}
\ddot{{x}^{i}}+{\delta}^{ik}n_{k,lj}{\dot{x}}^{l}{\dot{x}}^{j}=0
\label{5}
\end{equation}
where ${\Gamma}^{i}_{jk}=\frac{1}{2}{\delta}^{il}[g_{lj,k}+g_{lk,j}-g_{jk,l}]$
and we use the Euclidean 3D metric ${\delta}_{ij}$ to raise and lower indices.Let us now apply these ideas to the pure twist geometry of the nematic liquid crystals,where the director field is now given by the components
\begin{equation}
n_{z}=cos{\theta}(y)
\label{6}
\end{equation}
and
\begin{equation}
n_{x}=sin{\theta}(y)
\label{7}
\end{equation}
Where ${\theta}$ is the twist angle and the planar crystal is orthogonal to the y-coordinate direction.Substituting expressions (\ref{6}) and (\ref{7}) into (\ref{3}) one obtains the following components of the torsion vector $T^{i}_{ki}=T_{k}$
\begin{equation}
T_{y}({\theta})=-{{\partial}_{y}}(div n)
\label{8}
\end{equation}
and
\begin{equation}
T_{z}({\theta})=-{\nabla}^{2}n_{z}
\label{9}
\end{equation}
Since $div n=n_{x,x}+n_{y,y}+n_{z,z}=0$ , there is no torsion vector component along the orthogonal direction as happens in some domain walls with torsion \cite{11}.Thus the only non-vanishing torsion component of torsion reads
\begin{equation}
T_{z}({\theta})=-[cos{\theta}({\frac{d{\theta}}{dy}})^{2}+sin{\theta}\frac{d^{2}{\theta}}{{dy}^{2}}]
\label{10}
\end{equation}
Since local equilibrium conditions on the Liquid crystals yields \cite{10}
\begin{equation}
{\frac{d{\theta}}{dy}}=constant=K
\label{11}
\end{equation}
Thus equation (\ref{11}) reduces to 
\begin{equation}
T_{z}({\theta})=-K^{2}cos{\theta}
\label{12}
\end{equation}
This last expression tell us that the twisted geometry of the crystal leads 
to a 3D helical torsion.Geodesics of the twisted geometry of the liquid crystal leads to the results
\begin{equation}
\ddot{x}+n^{x}_{,yy}{\dot{y}}^{2}=0
\label{13}
\end{equation}
and 
\begin{equation}
\ddot{y}=0
\label{14}
\end{equation}
and finally
\begin{equation}
\ddot{z}+n^{z}_{,yy}{\dot{y}}^{2}=0
\label{15}
\end{equation}
Here the dots means derivation with respect to time coordinate.Substitution of equation (\ref{14}) into expressions (\ref{13}) and (\ref{15}) yields yields
\begin{equation}
\ddot{x}+{K_{1}}^{2}sin{\theta}(y)=0
\label{16}
\end{equation}
and
\begin{equation}
\ddot{z}+{K_{1}}^{2}cos{\theta}(y)=0
\label{17}
\end{equation}
To simplify matters let us solve these euqtions in the approximation of small twist angles ${\theta}<<<0$ where the equations (\ref{16}) and (\ref{17}) reduces to 
\begin{equation}
\ddot{x}+{K_{1}}^{2}{\theta}=0
\label{18}
\end{equation}
and
\begin{equation}
\ddot{z}+{K_{1}}^{2}=0
\label{19}
\end{equation}
Therefore from expression (\ref{21}) one obtains the following solution
\begin{equation}
z=-{K_{1}}^{2}t+c
\label{20}
\end{equation}
where c is an integration constant.To solve the remaining equations we need an explicit form of the twist angle with respect to time ,this can be obtained from the integration of the expression ${\frac{d{\theta}}{dy}}=0$
which yields  
\begin{equation}
{\theta}={K}y+f
\label{21}
\end{equation}
where $f$ is another integration constant.Substitution of $y=K_{0}t+d$ into this equation yields
\begin{equation}
{\theta}=mt+g
\label{22}
\end{equation}
where $f$,$m$ and $g$ are new integration constants.Substitution of (\ref{22}) into (\ref{18}) yields
\begin{equation}
\ddot{x}=-({K_{1}}^{2}mt+K_{1}g)
\label{23}
\end{equation}
Integration of this expression yields
\begin{equation}
x(t)=-\frac{1}{6}({\alpha}t^{3}+{\beta}t^{2}+{\gamma}t+{\delta})
\label{24}
\end{equation}
substitution of the expression $t=\frac{y}{K_{0}}$ into (\ref{24}) yields
\begin{equation}
x=-\frac{1}{6}({\alpha}^{'}y^{3}+{\beta}^{'}y^{2}+{\gamma}^{'}y+{\delta})
\label{25}
\end{equation}
where all the greek letters represent new integration constants.Therefore the trajectory of the 
test particles is represented by a third order polinomial curve.Trajectories of test particles 
in domain walls are in general parabolic curves.After we finish this letter we hear that 
Dubois-Violette and Pansu \cite{11} have considered a similar application of teleparallelism to 
cholesteric Blue Phase of liquid crystals.Nevertheless in their paper Cartan torsion is constant 
and thus our approach seems to be more general. 
\section*{Acknowledgement}
I am very much indebt to Prof.P.S.Letelier and Prof.F.Moraes  for helpful discussions on the 
subject of this paper.Financial support from CNPq. and UERJ is gratefully acknowledged.


\begin{thebibliography}{11}
\bibitem{1}M.O.Katanaev and I.Volovich,Annals of Physics,(1990).
\bibitem{2}S.Amari,(1962)RAAG memoirs 3,257.
\bibitem{3}L.C.Garcia de Andrade,On Riemannian 3D geometry of Heisenberg ferromagnets,gr-qc. 
\bibitem{4}A.Vilenkin and P.Shellard,(1995)Cosmic strings and other topological defects,Cambridge University Press.
\bibitem{5}L.C.Garcia de Andrade,(1999)Bloch walls with spin density as planar topological defects.
\bibitem{6}L.C.Garcia de Andrade,(1998)J.Math.Phys.,39.
\bibitem{7}Mullen,Geometrical defects in Josephson junction arrays,(1999).
\bibitem{8}F.Moraes and C.Furtado,Europhysics Lettters(1999). 
\bibitem{9}A.Einstein,(1928)Berliner Sitzungsber,217. 
\bibitem{10}P. de Gennes and ,(1990)The Physics of Liquid Crystals,Oxford University Press. 
\bibitem{11}E.Dubois-Violette and E.Pansu in Geometry of condensed matter physics,(1990)
Directions in condensed matter physics,vol.9.
\end{thebibliography}
\end{document}